\documentclass[aps,prb,twocolumn,amsmath,amssymb]{revtex4}
\usepackage{times}
\usepackage{graphicx}
\usepackage{amsfonts}
\usepackage{amsmath, amsthm, amssymb}
\usepackage{dsfont}
\usepackage{color}
\usepackage{sidecap}

\newcommand{\ve}[1]{\boldsymbol{#1}}
\DeclareMathOperator{\diag}{diag} 

\newcommand{\vecA}{\ve{A}} 

\newcommand{\hatz}{\hat{\boldsymbol{z}}} 

\newcommand{\ie}{\textit{i.e. }}

\usepackage{soul}

\begin{document}

\title{Quantum kinetic equations and anomalous non-equilibrium Cooper pair spin accumulation in Rashba wires with Zeeman splitting} 

\author{Sol H. Jacobsen and Jacob Linder}

\affiliation{Department of Physics, NTNU Norwegian University of
Science and Technology, N-7491 Trondheim, Norway}
\affiliation{Center for Quantum Spintronics, Department of Physics, Norwegian University of Science and Technology, NO-7491 Trondheim, Norway}

\begin{abstract}
We derive the theoretical and numerical framework for investigating nonequilibrium properties of spin-orbit coupled wires with Zeeman splitting proximized by a superconductor in the non-linear diffusive regime. We demonstrate that the anisotropic behaviour of triplet Cooper pairs in this system leads to novel spin accumulation profiles tunable by the magnetic field and strength of applied voltage bias. This paves the way for enhanced manipulation of superconducting spintronic devices, and enables further investigation of nonequilibrium effects in proximity-coupled superconducting structures more generally.
\end{abstract}

\date{\today}

\maketitle

\section{Introduction}
Superconducting spintronics has captured the minds of theoreticians and experimentalists alike with the promise of low-dissipation control of charge and spin transport in cryogenic devices \cite{linder_nphys_15, eschrig_physrep_15}.  This wave of interest has resulted in new insights into the underlying physical mechanisms, such as those that generate and control long-range spin-polarized triplet supercurrents \cite{bergeret_prl_01, eschrig_prl_09, alidoust_prb_10, shomali_njp_11, moor, halterman, BT13, BT14, JOL2015, JacobsenLinder2015, JKL2016}. This is accompanied by encouraging experimental results demonstrating key features such as enhanced quasiparticle spin lifetimes \cite{yang_nmat_10}, spin relaxation lengths \cite{quay_nphys_13}, spin Hall effects \cite{wakamura_nmat_15} and the confirmation of long-range dissipationless current through both strong ferromagnets \cite{keizer_nature_06, anwar_prb_10, khaire_prl_10, robinson_science_10} and spin-polarized Cooper pairs induced in conventional superconductors \cite{dibernardo_natcom_15, kalcheim_prb_15}.

In this work, we derive the quantum kinetic equations for investigating nonequilibrium properties of spin-orbit coupled nanowires with Zeeman splitting proximized by a conventional $s$-wave superconductor, and investigate these both numerically and analytically.  As an application of the equations, we investigate spin accumulation in such a nanowire under a voltage bias. We study the regime where the exchange field is of the same order of magnitude as the superconducting gap $\Delta$, 
which may be induced via an external magnetic field, or included via an intrinsically ferromagnetic wire. Spin accumulation effects in superconducting systems without spin-orbit coupling have been studied in previous works \cite{jedema_prb_99, sengupta_prl_08, lu_prb_09, shevtsov_prb_14, bobkova_jetp_15}. When the exchange splitting is much greater than $\Delta$ the superconducting proximity effect is negligible and the transport properties are instead governed by the interface~\cite{Belzig2000}. We demonstrate that intrinsic spin-orbit coupling (SOC), which has been shown to be instrumental in generating and controlling equilibrium charge and spin supercurrents in homogeneous ferromagnets \cite{BT13, BT14, JKL2016}, reveals novel features in the spin accumulation due to the anisotropic behaviour of triplet Cooper pairs. We show that the spin accumulation perpendicular to the field can be switched on and off, and that the spin accumulation may oscillate within the sample as a linear function of the strength of the SOC. Moreover, a small increase in the field strength may either shift the maximal magnetization towards higher bias values, or enhance the peak magnetization without a bias-shift, depending on the field rotation. To supplement the analytical considerations of the effects mentioned above, we also briefly examine the local density of states and charge conductance of the system.

\section{Theory} 
\begin{figure}
\includegraphics[width=0.48\textwidth, angle=0,clip]{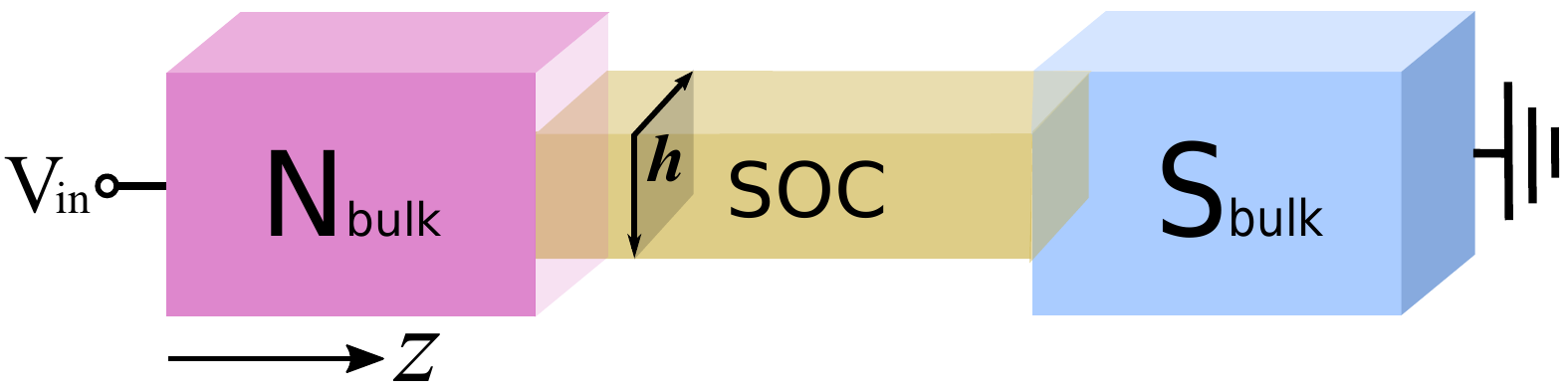}
\caption{(Color online) The model: nanowire with intrinsic spin-orbit coupling and Zeeman splitting, proximized by an $s$-wave superconductor and a voltage bias $V_{\rm{in}}$ applied via a bulk normal metal. The magnetization exchange field $\boldsymbol{h}$ is purely in-plane.}
\label{fig:model}
\end{figure}

When materials have intrinsic or extrinsic SOC, the injection of a charge current leads to transverse spin accumulation, \ie induced nonequilibrium magnetization, along the sample edges, known as the spin Hall effect \cite{DP_1_1971, DP_2_1971,SHRRev2015}. We will consider the effective one-dimensional (1D) heterostructure depicted in Fig.~\ref{fig:model}, which shows a nanowire with intrinsic SOC and Zeeman splitting. The wire is proximized by a conventional $s$-wave superconductor, and a voltage bias is applied to the system via a bulk normal metal. Here, we consider a small magnetization exchange field oriented in the plane of the cross-section (hereafter referred to as in-plane), and this field may be either intrinsic or extrinsic.  The latter case can be achieved via an external field or a proximate ferromagnetic insulator. The case of an externally applied field should be particularly suitable for producing an in-plane exchange field, whereas an intrinsically occurring magnetization likely would favor a field oriented along the wire due to shape anistoropy (we briefly discuss this case in the Appendix, for completeness). Since the nanowire is considered to be a 1D structure, the spin accumulation does not occur on any lateral faces but varies along the wire.

The appearance of the long-range (LR) spin-polarized triplet component of the superconducting correlations occurs in the presence of SOC when the Cooper pair spins have projections parallel to the exchange field. The components of the spins with perpendicular projection relax over much shorter distances of the order of the wire's coherence length. Formally, the criterion for the existence of the LR component in a nanowire with the spin-orbit gauge field $\boldsymbol{A}$ oriented along the wire is that the commutator $\left[\boldsymbol{A},\boldsymbol{h}\cdot\boldsymbol{\sigma}\right]$ must be nonvanishing \cite{BT14}, where $\boldsymbol{h}$ is the exchange field vector and $\boldsymbol{\sigma}$ is the Pauli vector \footnote{Note that whenever the axis of inversion symmetry breaking causing intrinsic SOC is not perpendicular to the layering direction of heterostructures the criterion for the existence of the LR triplet component is $\left[\boldsymbol{A},\left[\boldsymbol{A},\boldsymbol{h}\cdot\boldsymbol{\sigma}\right]\right]$ (see \cite{BT14}).}$^,$\footnote{Note that the term ``spin orbit field'' is sometimes used in the literature to denote the quantity $B$ that enters the Hamiltonian as $B\cdot\boldsymbol{\sigma}$. In contrast our SOC enters the Hamiltonian as $A\cdot \boldsymbol{p}$, and hence we use the term ``SO gauge field'' to avoid ambiguity.}. However, satisfying the criterion does not guarantee experimentally measurable observables or features useful for experimental manipulation of the system \cite{JOL2015,JacobsenLinder2015}. By deriving the quantum kinetic equations for driven, diffusive superconductor-ferromagnet systems, we can investigate the nonequilibrium dynamics and extract experimentally measurable markers. Below we will demonstrate that the anisotropy of the triplet Cooper pairs leads to novel control over the spin accumulation in such nonequilibrium structures, tunable via the relationship between the exchange field and the external bias.

To derive the quantum kinetic equations and investigate the diffusive regime of the above heterostructure out of equilibrium we employ the Keldysh formulation of quasiclassical theory \cite{Belzig1999, Chandrasekhar2004}, in which the full $8\times 8$ Green's function $\check{g}$ is expressed in terms of its retarded (R), advanced (A) and Keldysh (K) components as
\begin{eqnarray}
	\check{g}=\begin{pmatrix}
	\hat{g}^R & \hat{g}^K\\ \hat{0} & \hat{g}^A 
	\end{pmatrix}.
\end{eqnarray}
In the absence of SO coupling, the Usadel equation \cite{Usadel1970} for the nanowire reads
\begin{eqnarray}\label{Eqn:Usadel1}
	D_F\nabla(\hat{g}\nabla \hat{g})+i\left[\epsilon\hat{\rho}_3+\rm{diag}\it(\boldsymbol{h}\cdot\boldsymbol{\sigma},(\boldsymbol{h}\cdot\boldsymbol{\sigma})^*), \hat{g}\right]=0,
\end{eqnarray}
where $\epsilon$ denotes the quasiparticle energy, $^*$ denotes complex conjugation, and the matrix $\hat{\rho}_3=\rm{diag}(1,1,-1,-1)$. To investigate the Keldysh component of the Green's function explicitly, we take $\hat{g}^K=\hat{g}^R \hat{h}-\hat{h}\hat{g}^A$, and substitute this into Eq.(\ref{Eqn:Usadel1}). Here the matrix $\hat{h}$ contains the distribution functions, and the advanced component of the Green's function is simply $\hat{g}^A=-\hat{\rho_3}\hat{g}^{R\dagger}\hat{\rho_3}$.

The SOC is included in the model by replacing all derivatives with their gauge covariant counterpart \cite{BT14,Gorini2010}
\begin{eqnarray}
	\nabla(\,\cdot\,) \mapsto \tilde{\nabla}(\,\cdot\,) \equiv \nabla(\,\cdot\,) -i\left[\boldsymbol{\hat{A}},\;\cdot\;\right] \, ,
\end{eqnarray}
where $\boldsymbol{\hat{A}}$ has both a vector structure in geometric space, and a $4\times 4$ matrix structure in spin--Nambu space: $\boldsymbol{\hat{A}} = \mathrm{diag}(\boldsymbol{A},-\boldsymbol{A}^*)$, with the SO gauge field $\boldsymbol{A} = (A_x,A_y,A_z)$. For the case of a nanowire with pure Rashba SOC as in Fig.~\ref{fig:model}, the only non-zero component of $\boldsymbol{A}$ is $A_z = \alpha(\sigma_x-\sigma_y)$, where $\sigma_i$ denote the usual Pauli matrices. The coefficient $\alpha$ is normalized to the superconducting gap $\Delta$ and sample length $L$, such that for $\Delta\approx 1-3$~meV and $L/\xi_S=0.8$, the SOC will be of the order $\alpha\approx 10^{-11}-10^{-12}$~eV m, which agrees well with experimental estimates \cite{manchon_natmat_15}.

The Riccati-parametrized Usadel equation for $\hat{g}^{R}$ including intrinsic SOC was derived for the equilibrium case in Ref.\onlinecite{JOL2015}, and is provided in Appendix~\ref{App:Riccati}. The remaining step in order to investigate the Keldysh component explicitly is then to find the corresponding equation for the distribution functions:
\begin{widetext}
\begin{align}\label{Eqn:h}
D_F\left(\partial_z^2\hat{h}-(\hat{g}^R(\partial_z^2\hat{h})\hat{g}^A\right) &=   +i\left[\hat{g}^K,\epsilon\hat{\rho}_3+\rm{diag}\it(\boldsymbol{h}\cdot\boldsymbol{\sigma},(\boldsymbol{h}\cdot\boldsymbol{\sigma})^*)\right] + D_F\Big(-(\hat{g}^R\partial_z\hat{g}^R)(\partial_z\hat{h})+(\partial_z\hat{h})(\hat{g}^A\partial_z\hat{g}^A) \notag\\
& + (\partial_z\hat{g}^R)(\partial_z\hat{h})\hat{g}^A + \hat{g}^R(\partial_z\hat{h})(\partial_z\hat{g}^A)  + i\hat{g}^K\hat{A}_z(\partial_z\hat{g}^A) + i\hat{A}_z\hat{g}^K(\partial_z\hat{g}^A) - 2i\hat{g}^R(\partial_z\hat{g}^K)\hat{A}_z  \notag\\
 &- 2i\hat{g}^K(\partial_z\hat{g}^A)\hat{A}_z + i(\partial_z\hat{g}^R)\hat{A}_z\hat{g}^K - i(\partial_z\hat{g}^R)\hat{g}^K\hat{A}_z + i\hat{g}^R\hat{A}_z(\partial_z\hat{g}^K) + i\hat{A}_z\hat{g}^R(\partial_z\hat{g}^K)  \notag\\
&+ i(\partial_z\hat{g}^K)\hat{A}_z\hat{g}^A - i(\partial_z\hat{g}^K)\hat{g}^A \hat{A}_z+ \boldsymbol{\hat{A}}\it\hat{g}^R\boldsymbol{\hat{A}}\it\hat{g}^K - \hat{g}^R\boldsymbol{\hat{A}}\it\hat{g}^K\boldsymbol{\hat{A}}\it + \boldsymbol{\hat{A}}\it\hat{g}^K\boldsymbol{\hat{A}}\it\hat{g}^A - \hat{g}^K\boldsymbol{\hat{A}}\it\hat{g}^A\boldsymbol{\hat{A}}\it  \notag\\
&- (\partial_z\hat{g}^R\partial_z\hat{g}^R)\hat{h} - \hat{g}^R(\partial_z^{\rm{2}}\it\hat{g}^R)\hat{h} + \hat{h}(\partial_z\hat{g}^A\partial_z\hat{g}^A) + \hat{h}\hat{g}^A(\partial_z^{\rm{2}}\it\hat{g}^A)\Big).
\end{align}
\end{widetext}
Here $\hat{A}_z=\text{diag}(A_z,-A_z^*)$ is the only non-zero component of the field $\boldsymbol{\hat{A}}\rm$ in the special case of a nanowire oriented along the junction, where $A_x=A_y=0$. The corresponding Kupriyanov-Lukichev boundary conditions\cite{KuprianovLukichev1988} take the form
\begin{align}\label{Eqn:KL}
2L_j&\zeta_j\left(\partial_z\hat{h}_j -\hat{g}^R_j(\partial_z\hat{h}_j)\hat{g}^A_j\right) = [\check{g}_L,\check{g}_R]^K \notag\\
&- 2L_j\zeta_j\Big(\hat{g}^R_j(\partial_z\hat{g}^R_j)\hat{h}_j-\hat{h}_j\hat{g}^A(\partial_z\hat{g}^A_j) - i\hat{g}^R_j\hat{A}_z\hat{g}^R\hat{h} \notag\\
&+ i\hat{h}\hat{g}^A\hat{A}_z\hat{g}^A + i\hat{g}^R\hat{A}_z\hat{h}\hat{g}^A - i\hat{g}^R\hat{h}\hat{A}_z\hat{g}^A\Big),
\end{align}
where subscripts $j=\left\{L,R\right\}$ indicate the left and right sides of the interface. Finding $\hat{g}^K$ in practice then involves solving a series of coupled partial differential equations for both $\hat{g}^R$ and $\hat{h}$. This can be achieved numerically by first employing the Riccati parametrization \cite{Schopohl1995,JOL2015} to solve for $\hat{g}^R$ (see Appendix for details), and using this as an input in solving for $\hat{h}$ via Eq.~(\ref{Eqn:h}), as we do below.

Once the Keldysh Green's function is found, the spin accumulation $M_\tau$  along the unit vector $\hat{\tau}$ is found as follows \cite{ChampelEschrig2005,Alexander1985}:
\begin{align}
M_{\tau} &= M_{0} \int^\infty_{-\infty} \text{d}\epsilon \text{Tr}\{ \hat{\tau}\diag(\boldsymbol{\sigma},\boldsymbol{\sigma}^*) \hat{g}^K \}.\label{eq:spinacc}
\end{align}
The constant $M_0=g\mu_B N_0\Delta/16$, with the Land\'e $g$-factor $g\approx 2$ for electrons and $\mu_B$ is the Bohr magneton. The unit vector $\hat{\tau}$ determines which polarization component of the spin accumulation that is investigated. For instance, computing the component of the spin accumulation along the exchange field $\boldsymbol{h}$ renders $\hat{\tau}$ equal to the unit vector of the exchange field. Similarly, one can define vectors perpendicular to the field to probe the spin accumulation polarization perpendicular to $\boldsymbol{h}$. We underline that Eq. (\ref{eq:spinacc}), which we will refer to as the total spin accumulation when there may be ambiguity, contains an equilibrium and non-equilibrium contribution. The equilibrium contribution exists even in the absence of any applied voltage and describes the proximity-induced magnetization due to the presence of odd-frequency triplet Cooper pairs coexisting in a non-unitary fashion with singlet pairs. The non-equilibrium contribution (often by itself called spin accumulation in the literature) exists only in the presence of an applied voltage. Different measurement methods may be employed to measure either the total or non-equilibrium components, and therefore we shall present results both for the total spin accumulation and the pure non-equilibrium part below.

Herein we will also examine the normalized charge conductance $\sigma/\sigma_\infty$, where $\sigma=\delta I_Q/\delta V$ and $\sigma_\infty$ is the normal-state value at high bias, $eV\gg\Delta$, $V$ is the applied voltage bias in Volts, $e$ is the electronic charge, and the charge current $I_Q$ is found by:
\begin{align}
I_{Q} &= I_{Q_0} \int^\infty_{-\infty} \text{Tr}\{ \hat{\rho}_3 (\check{g}\partial_z\check{g})^K \}\text{d}\epsilon,\label{eq:chargecur}
\end{align}
Here $I_{Q_0}=N_0 D A \Delta e/4L$, where $N_0$ is the normal-state density of states at the Fermi level, $D$ is the diffusion constant and $A$ the interfacial contact area. The integral in Eqs.~(\ref{eq:chargecur}) is dimensionless since the energies have been normalized to the bulk superconducting gap $\Delta$ and lengths normalized to the nanowire length $L$. When including SOC, the explicit expression for charge current becomes:
\begin{eqnarray}
	I_Q = I_{Q_0}\!\!\int^\infty_{-\infty}\!\! \text{Tr}\{ \rho_3(\partial \hat{h} + \hat{g}^R\partial \hat{g}^R \hat{h} - \hat{h}\hat{g}^A\partial\hat{g}^A - \hat{g}^R\partial\hat{h}\hat{g}^A \nonumber\\
	 - i\hat{g}^R \boldsymbol{\hat{A}}\it\hat{g}^R\hat{h} + i\hat{h}\hat{g}^A \boldsymbol{\hat{A}}\it\hat{g}^A + i\hat{g}^R \boldsymbol{\hat{A}}\it\hat{h}\hat{g}^A - i\hat{g}^R \hat{h}\boldsymbol{\hat{A}}\it\hat{g}^A  )\}\text{d}\epsilon. \nonumber\\
\end{eqnarray}

\section{Results}
Having derived the kinetic equations for our nonequilibrium system, we can use this result to consider the effect of voltage bias for a Zeeman-split nanowire with intrinsic SOC as in Fig.~\ref{fig:model}. To ensure the superconducting proximity effect is dominant when we apply a voltage bias, we set the interface parameter (ratio of bulk-to-interfacial resistance) $\zeta_1=15$ at the normal interface, corresponding to the strong tunneling limit, and $\zeta_2=3$ at the superconductor interface. We set the temperature $T=0.005T_C$. All spin accumulation amplitudes are given in units of $2M_0$ and the numerical integration is in practice performed over the range $0-\Omega$, where $\Omega$ is a suitable high-energy cutoff. Taking a normal-state density of states $N_0\sim 10^{22}$ eV$^{-1}$cm$^{-3}$, $\Delta\sim 1$ meV and $\mu_B=5.788\times 10^{-5}$ eVT$^{-1}$, the value of the normalization constant $M_0\approx 7.2\times 10^{13}$ eV/Tcm$^3$.

\begin{figure*}
\includegraphics[width=\textwidth, angle=0,clip]{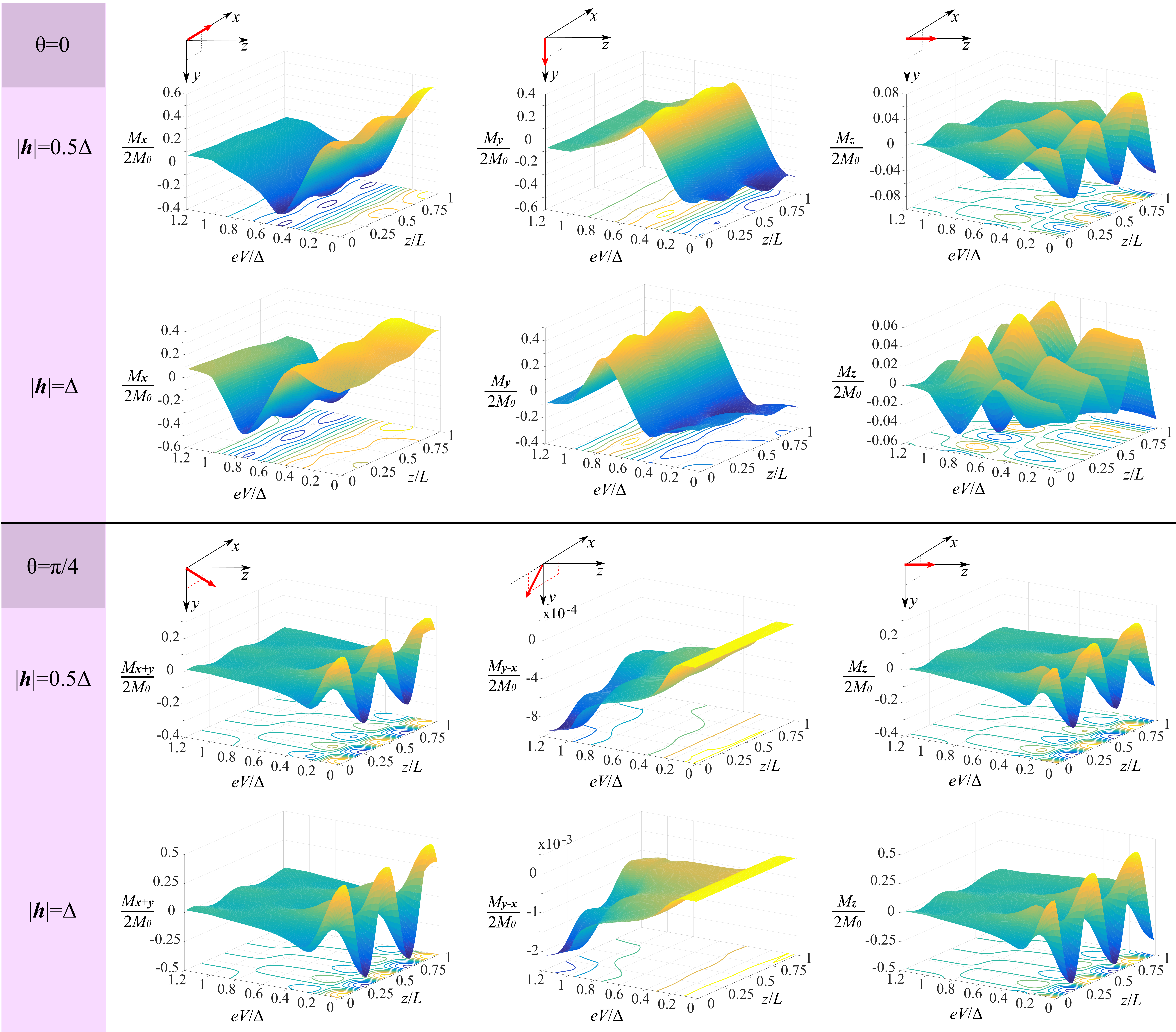}
\caption{(Color online) Total spin accumulation $M_\tau$ along unit vectors $\tau$ for a nanowire with SOC and Zeeman splitting as a function of the applied voltage bias. Two field orientations are shown: $\boldsymbol{h}=|\boldsymbol{h}|(1,0,0)$ in the upper half, and $\boldsymbol{h}=|\boldsymbol{h}|/\sqrt{2}(1,1,0)$ in the lower half. Two field strengths are displayed for each orientation: $|\boldsymbol{h}|=0.5\Delta$ in the first row, and $|\boldsymbol{h}|=\Delta$ in the second row. The first column represents spin accumulation along the field (as indicated on the coordinate schematic), while the others show perpendicular components. The nanowire has length $L/\xi_S=0.8$ and the SOC is of pure Rashba type with coefficient $\alpha=5/L$. Spin accumulation amplitudes are given in units of $2M_0$.}
\label{fig:SpinAcc}
\end{figure*}

In Fig.~\ref{fig:SpinAcc} we show how the total spin accumulation varies with applied bias along the length of the wire for the case $\alpha=5/L$. Two values of the exchange field strength $|\boldsymbol{h}|=0.5\Delta$ and $|\boldsymbol{h}|=\Delta$ are given, as well as two field orientations $\theta=0$ and $\theta=\pi/4$, where the exchange field $\boldsymbol{h}=|\boldsymbol{h}|(\cos\theta,\sin\theta,0)$. We remind the reader that the spin-orbit gauge field $\vecA$ points along the wire ($\vecA \parallel \hatz$), and thus the exchange field is always perpendicular to $\vecA$. For comparison, we also provide in Appendix~\ref{App:Noneq_fieldeff} the corresponding plots of the spin accumulation without the equilibrium component, \ie replacing $\hat{g}^K$ in Eq.(\ref{eq:spinacc}) by $\hat{g}^K-\hat{g}^K_{eq}$, where $\hat{g}^K_{eq}=(\hat{g}^R-\hat{g}^A)\tanh(0.5\epsilon/k_BT)$ is the usual Keldysh component in equilibrium, $k_B$ is the Boltzmann constant and $T$ is temperature. 

By comparing Figs.~\ref{fig:SpinAcc} and \ref{fig:NonEqSpinAcc}, we see as expected that the equilibrium component dominates at low bias. In contrast, the equilibrium and nonequilibrium portions tend to cancel towards higher bias so that the total spin accumulation tends to zero as $eV>\Delta$. Consequently, there is a suppression in magnitude of the total spin accumulation at intermediate bias, but the interesting nonmonotonic behaviour of the off-set nonequilibrium portion as a function of the applied voltage $V$ survives and matches well to the behaviour of the total spin accumulation in this region. This can be physically understood from the fact that the equilibrium component of the spin accumulation is completely independent of the applied voltage for a fixed value of $L$. On the other hand, both the equilibrium and non-equilibrium contributions to Eq.~(\ref{eq:spinacc}) vary with the length $L$ of the system as shown. 

Consider now how the total spin accumulation varies within the sample by increasing the exchange field $|\boldsymbol{h}|=0.5\Delta\rightarrow\Delta$ in Fig.~\ref{fig:SpinAcc}. When $\alpha=0$ (not shown), the spin accumulation exists along the field only, and rotating the field has no effect. In this case, increasing the field strength has the effect of shifting the peak magnetization towards higher bias values. With SOC, increasing the field strength induces this shift in the peak magnetization only for field orientation $\theta=0$, something which is also seen in Fig.~\ref{fig:NonEqSpinAcc}. By contrast, the spin accumulation for $\theta=\pi/4$ is not bias-shifted, but the magnitude of magnetization instead increases. 

Secondly, we note the effect of the exchange field rotation for the case $|\boldsymbol{h}|=0.5\Delta$ (rows 1 and 3 of Fig.~\ref{fig:SpinAcc}), where several interesting features appear. We see that the oscillation period, determined by the wavevector of the anomalous Green's function, does not vary appreciably between the cases. By contrast, the amplitude of oscillation varies greatly with applied voltage bias, in particular for the spin accumulation along the field direction. Moreover, we see that we can in effect \emph{turn off} the spin accumulation perpendicular to the field by rotating the field away from pure $\boldsymbol{\hat{x}}$-alignment. To explain all these features of the spin accumulation, we will now consider the underlying numerical structure and analytic limits in more detail.

We examine the numerical solution to the Usadel equations in the Riccati formalism (Eqn.~(\ref{Eqn:SOUsadel})) using $f=2N\gamma$ for the anomalous Green's function (top right quadrant of $\hat{g}^R$ in Eqn.~(\ref{Eqn:gR})). We re-express $f$ in the so-called $d$-vector formalism, in which $f=(f_s+\boldsymbol{d}\cdot\boldsymbol{\sigma})i\sigma_y$. In that case the short-range triplets are those aligned with the exchange field $d_\parallel=\boldsymbol{d}\cdot\boldsymbol{\hat{h}}$, and the long-range triplets are perpendicular, \ie $d_\perp=|\boldsymbol{d}\times\boldsymbol{h}|$. In that case we may plot the components of the $d$-vector, as we have shown in Fig.~\ref{fig:dvec} in the Appendix. Firstly, we see that the triplet components reflect the dominance of the equilibrium component of the magnetization at low bias, which was evident from Fig.~\ref{fig:NonEqSpinAcc}, as well as the diminution of total spin accumulation at high bias. Secondly, we can see qualitatively that the oscillations in the spin accumulation follow directly from the real part of the corresponding components of the anomalous Green's function. That is, for $\theta=0$, the spin accumulation along the field, $M_x$, follows $d_x=d_\parallel$, the first perpendicular component, $M_y$, follows $d_y=d_\perp$, and the second perpendicular, $M_z$, follows $d_z$. For $\theta=\pi/4$, we see in Fig.~\ref{fig:dvec} that $d_x=d_y$. Thus we have an enhanced component of spin accumulation along the field $M_{x+y}$, while the first perpendicular component $M_{y-x}$ is significantly diminished -- indeed it is zero when the bias is negligible -- since the dominant component of the anomalous Green's function is $d_\perp=(d_x-d_y)/\sqrt{2}$. That the spin accumulation should be closely related to the $d$-vector is reasonable, since we know that in equilibrium the induced magnetization can be written as a Matsubara-sum over the product of $f_s$ and the triplet vector\cite{ChampelEschrig2005}.

We may examine the zero-bias case analytically in the weak proximity limit. To derive the analytic solution to the non-equilibrium Usadel equations we employ the Riccati parameterization as given in Appendix~\ref{App:Riccati}, along with the fact that $|\gamma_{ij}|\ll 1$ and $N\approx 1$ in the weak proximity limit. Using the $d$-vector formalism, the weak proximity equations become:
\begin{eqnarray}\label{Eqn:WPE}
D_F\partial^2 f_s \!&=&\! -2i(\epsilon f_s+|\boldsymbol{h}| d_\parallel),\nonumber\\
D_F\partial^2d_\parallel \!&=&\! -2i\epsilon d_\parallel - 2i|\boldsymbol{h}|f_s + 4D_F\alpha^2(1-i\cos(2\theta))d_\parallel \nonumber\\
&& \!+ 4D_F\alpha^2i\sin(2\theta)d_\perp \nonumber\\
&& \!+ 4D_F\alpha\partial(d_z)(\cos\theta+\sin\theta),\nonumber\\
D_F\partial^2 d_\perp \!&=&\! -2i\epsilon d_\perp +4D_F\alpha^2(1+i\cos(2\theta))d_\perp \nonumber\\
&& \!+ 4D_F\alpha^2i\sin(2\theta)d_\parallel \nonumber\\
&& \!- 4D_F\alpha \partial(d_z)(\sin\theta-\cos\theta),\nonumber\\
D_F\partial^2 d_z \!&=&\! -2i\epsilon d_z + 8D_F\alpha^2d_z - 4D_F\alpha\partial(d_y+d_x).
\end{eqnarray}
From the equation for $d_z$ in (\ref{Eqn:WPE}), we see that the SO parameter introduces pair-breaking (damping) that scales with $\alpha^2$. Moreover, it is clear from all the equations that $\alpha$ introduces a coupling between all triplet $d$-vector components. The solutions to Eqns.~(\ref{Eqn:WPE}) are given explicitly in Appendix~\ref{App:WPE} for both field orientations $\theta=0$ and $\pi/4$, corresponding to those in Fig.~\ref{fig:SpinAcc}. By examining Eqns.~(\ref{Eqn:WPE}), we see that the long-range triplet component $d_\perp$ cannot be generated from $d_\parallel$ when $\theta=0$. Consistently, the solution for $d_\perp$ (given in Eqn.~(\ref{Eqn:0})) is independent of the exchange field for $\theta=0$. The solutions also allow us to extract the real and imaginary components of the wavevectors, which we will go on to use to gain insight into the damping and oscillation lengths.

\begin{figure*}
\includegraphics[width=\textwidth, angle=0,clip]{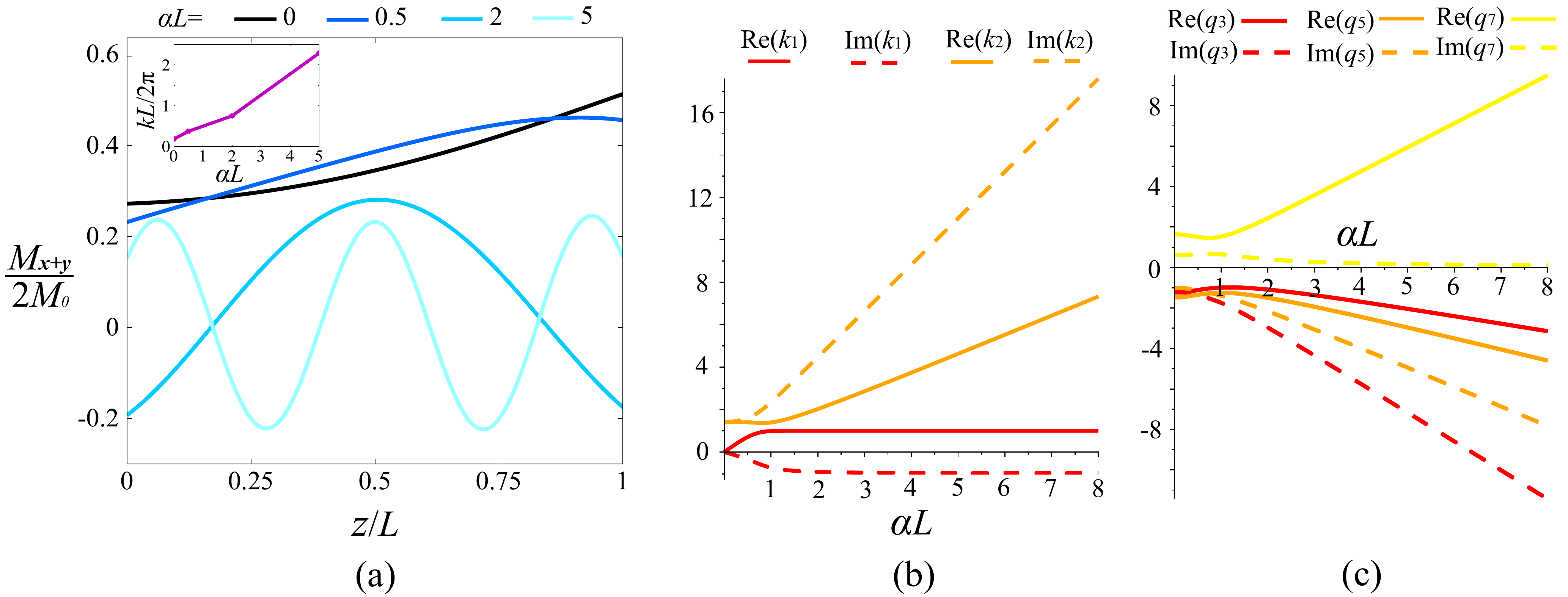}
\caption{(Color online) (a) Total spin accumulation along the field $\boldsymbol{h}=0.5\Delta/\sqrt{2}(1,1,0)$ for increasing SOC-strength at zero bias, in units of $2M_0$. The inset shows a linearly increasing spatial frequency of oscillation, derived by fitting the numerical curves to a sinusoid via least squares. Real and imaginary parts of the component wavevectors with increasing Rashba coupling $\alpha$ for the analytic weak proximity solution at field orientation $(b)$ $\theta=0$ (Eqns.~(\ref{Eqn:0})), and (c) $\theta=\pi/4$ (Eqns.~(\ref{Eqn:pi4})). The nanowire has length $L/\xi_S=0.8$.}
\label{fig:Reff}
\end{figure*}

In the case without spin-orbit coupling, the spin accumulation can only have a component along the field. Furthermore, increasing the field strength when there is no spin-orbit coupling is known to decrease the oscillation period of the triplet Green's functions~\cite{EschrigPhysTod2011}, which we can also reproduce for large field strengths (not shown). However, for weak fields we observe that the Rashba parameter $\alpha$ is by far the dominant factor governing the oscillation frequency of the spin accumulation within the sample. By examining Equations~(\ref{Eqn:0}) and (\ref{Eqn:pi4}), we see that the leading term in $\alpha$ in the wavevectors of the anomalous Green's functions are of first order for both field orientations, which explains why there is no appreciable difference in oscillation upon rotation in Fig.~\ref{fig:SpinAcc}. We show the total numerical spin accumulation at zero bias with increasing $\alpha$ in Fig.~\ref{fig:Reff}$(a)$, where the inset shows the respective spatial frequencies as a function of $\alpha$, derived by fitting the curves to a sinusoid via least squares. The analytical component wavevectors from the solution in the weak proximity regime as a function of $\alpha$ are given for field rotation $\theta=0$ in Fig.~\ref{fig:Reff}$(b)$, and for $\theta=\pi/4$ in \ref{fig:Reff}$(c)$. These plots confirm that the number of oscillations within the sample scales linearly with $\alpha$ both numerically and analytically for weak fields. 

In Figs.~\ref{fig:Reff}$(b)$ and $(c)$ we also see that both the damping and oscillation lengths, \ie the real and imaginary parts of the wavevector, become equal for $\alpha=0$ as expected. For the case $\theta=\pi/4$ this requires the component wavevectors (as defined in Appendix~\ref{App:WPE}) $q_5=q_7=0$ at $\alpha=0$, while they may be nonzero more generally. Notice also that the damping and oscillation lengths may diverge significantly with increasing $\alpha$, although one requires the solution to the equations for the boundary conditions in order to specify the relative proportion of each wavevector. It is interesting that the introduction of spin-orbit coupling may render the decay and oscillation lengths of the superconducting correlations to be very different.

\begin{figure}[b!]
\includegraphics[width=0.48\textwidth, angle=0,clip]{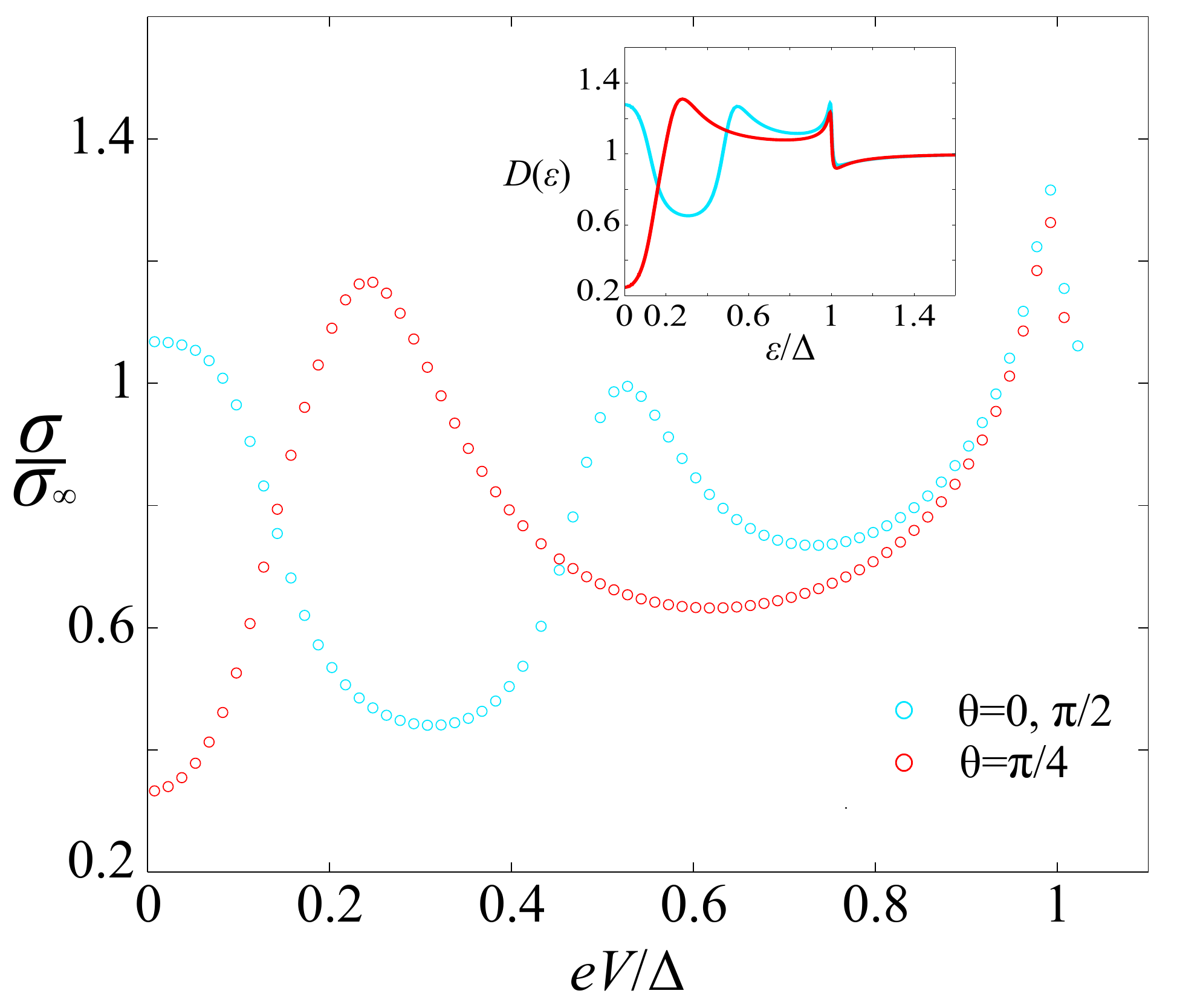}
\caption{(Color online) Normalized charge conductance as a function of applied voltage bias, for exchange field orientations $\theta=0$ and $\pi/4$, where $\boldsymbol{h}=|\boldsymbol{h}|(\cos\theta,\sin\theta,0)$. The nanowire has length $L/\xi_S=0.8$, exchange field strength $|\boldsymbol{h}|=0.5\Delta$, and the SOC is of pure Rashba type with coefficient $\alpha=5/L$. The inset shows the local density of states $D(\epsilon)$ in the middle of the sample.}
\label{fig:bias}
\end{figure}

To investigate the role of the triplet Cooper pairs further, we plot in Fig.~\ref{fig:bias} the normalized charge conductance as a function of the applied bias, for different orientations of the in-plane exchange field, and we note several features.
Firstly, we see that the charge conductance is symmetric about $\theta=\pi/4$ for the in-plane exchange field $\boldsymbol{h}=|\boldsymbol{h}|(\cos\theta,\sin\theta,0)$, and a quick calculation of the charge current shows that this is peaked at zero bias for $\theta=\left\{0,\pi/2\right\}$ and suppressed at $\theta=\pi/4$. This is consistent with the control of spectral features exhibited by ferromagnets with SOC, in which the local density of states can be tuned from fully gapped to peaked to normal-state at $\epsilon=0$ upon altering the orientation of the exchange field \cite{JOL2015}. For reference, the local density of states for the current example, which is not fully gapped but nevertheless displays a considerable peak-to-trough on rotation $\theta=0\rightarrow \pi/4$, is given in the inset of Fig.~\ref{fig:bias}. Note that the density of states is independent of applied voltage, but that the charge-conductance/bias profile follows that of the energy-distribution of the local density of states for the parameter regime considered here. The SOC-induced strong anisotropy in the conductance shown here is consistent with the strong anisotropies found in the supercurrent \cite{fabian_prl_15, costa_prb_17} and Meissner effect \cite{espedal_prl_16} in similar structures.

A gapped density of states indicates that the current is carried entirely by singlets, while a peak is the characteristic signature of long-range triplets. However, it has been shown for normal metals that the charge conductance can be peaked at low temperatures and zero bias simply due to coherent Andreev reflection when the proximity effect is strong~\cite{Yokoyama2005}. In this case the strength of the proximity effect is not altered, but the presence of SOC means the direction of the exchange field can be used to tune the carriers from singlets to triplets. By looking at the normal-state limit $\boldsymbol{h}=0$ (not shown), we can confirm that the case in which the carriers are predominantly singlet ($\theta=\pi/4$) corresponds to the superconductor-normal metal (SN) charge conductance profile (note that we can also recover the typical SN conductance profile for longer samples~\cite{Tanaka2003}).

Finally, we can now also explain why the peak in the spin accumulation is shifted to higher bias voltages upon increasing the exchange field, as shown in Figs.~\ref{fig:SpinAcc} and~\ref{fig:NonEqSpinAcc}. The shift towards higher bias values for $\theta=0$ can be explained by likening to a spin-split superconductor, in which the different spins of the spin-split minigaps increasingly move toward higher positive and negative energies in the density of states as the field is increased. Similarly, the bias would engender a shift in the magnetization, broadly following the profile of the density of states (see Fig.~\ref{fig:ldos} in the Appendix). Such energy-shifts in the spin-splitting have also been noted for increasingly spin-active interfaces, where the spin-active parameter takes the role of the exchange field\cite{Machon2014}. In contrast, Fig. \ref{fig:ldos} shows that the electronic spectral features remain very similar upon increasing the exchange field when $\theta=\pi/4$. As a result, the spin accumulation profile also remains qualitatively invariant for this orientation when $|\boldsymbol{h}|$ increases.

\section{Discussion and Outlook} 
We have derived the full quantum kinetic equations for studying nonequilibrium effects in diffusive heterostructures with intrinsic SOC, and provided an initial investigation of the effect of the exchange field and voltage bias on the spin accumulation and charge conductance in a nanowire with Zeeman splitting. The work concurs with previous investigations on charge conductance in SF systems, and crucially demonstrates how SOC facilitates the tuning of the spin accumulation via the strength and angle of the exchange field, SOC-strength, voltage bias, and position along the wire. We have shown that the spin accumulation perpendicular to the field can be eliminated, and we have shown both numerically and analytically that the spin accumulation oscillates within the sample as a linear function of the strength of the SOC. We have seen that a small increase in the field strength shifts the maximal magnetization towards higher bias voltages for certain orientations of the field. For other directions, the spin accumulation profile remains qualitatively invariant as $|\boldsymbol{h}|$ is varied in a regime of order $\sim \Delta$. This anisotropic behavior is a direct consequence of how the relative orientation of the exchange field and the spin-orbit vector causes formation of triplet Cooper pairs. 

More generally, the analytic framework for investigating the nonequilibrium physics of diffusive heterostructures with superconducting elements that we present here opens up a vast range of new phenomena to be explored, and we anticipate many new discoveries to be made in this direction in the near future. A natural next step would be to include the recently derived generalized boundary conditions for arbitrarily strong spin-polarization at interfaces \cite{eschrig_njp_15} to include spin-active heterostructures. Although a component of the SOC along the junction direction is required to affect the boundary conditions, it would also be instructive to look at thin-film examples where Dresselhaus coupling could be included, such as in InAs, since this has been shown to enhance control by some orders of magnitude~\cite{JOL2015,JKL2016}. Finally, it will be interesting to compare the present results with a complementary analysis of the effect of SOC in wires or ferromagnets with a thermal gradient at zero voltage bias, before considering the full case of simultaneous voltage and temperature difference in the system.

\textit{Acknowledgments}. We thank J.A. Ouassou for useful discussions on computational aspects. This research was supported in part with computational resources at NTNU provided by NOTUR, http://www.sigma2.no, and we acknowledge funding via the ``Outstanding Academic Fellows'' programme at NTNU, the COST Action MP-1201, the Research Council of Norway Grant numbers 216700 and 240806, as well as the Research Council of Norway's funding for the QuSpin Center of Excellence in Quantum Spintronics. \\

\appendix

\begin{widetext}
\section{Appendix}
\subsection{Riccati parameterization}\label{App:Riccati}
The Riccati parametrization for $\hat{g}^R$ is given by
\begin{eqnarray}\label{Eqn:gR}
	\hat{g}^R=
	\begin{pmatrix}
	N(1+\gamma\tilde{\gamma}) & 2N\gamma\\
	-2\tilde{N}\tilde{\gamma} & -\tilde{N}(1+\tilde{\gamma}\gamma)
	\end{pmatrix},
\end{eqnarray}
where the normalisation matrix is $N=(1-\gamma\tilde{\gamma})^{-1}$ and the tilde operation denotes a combination of complex conjugation $i\rightarrow -i$ and energy $\epsilon\rightarrow -\epsilon$, with $\gamma\rightarrow \tilde{\gamma}$, $N\rightarrow \tilde{N}$. The Riccati-parametrized Usadel equation including intrinsic SOC can be solved independently of the solution for $\hat{h}$ and was derived for the equilibrium case in Ref.\onlinecite{JOL2015}, taking the form
\begin{eqnarray}\label{Eqn:SOUsadel}
	D_F\left(\partial_z^2 \gamma + 2(\partial_z \gamma)\tilde{N}\tilde{\gamma}(\partial_z \gamma)\right) &=& -2i\epsilon\gamma - i\boldsymbol{h} \cdot (\boldsymbol{\sigma}\gamma-\gamma\boldsymbol{\sigma}^*)\nonumber\\
	&&\,+D_F\left[\boldsymbol{AA}\it\gamma-\gamma\boldsymbol{A^*A^*}\rm+2(\boldsymbol{A}\it\gamma+\gamma\boldsymbol{A}^*\it)\tilde{N}(\boldsymbol{A}\it^*+\tilde{\gamma}\boldsymbol{A}\it\gamma)\right]\nonumber\\
	&&\,+2iD_F\left[(\partial_z \gamma)\tilde{N}({A}^*_z+\tilde{\gamma}{A}_z\gamma)+({A}_z+\gamma {A}^*_z\tilde{\gamma})N(\partial_z \gamma)\right]\!.
\end{eqnarray}
Here $\epsilon$ is the quasiparticle energy, $\boldsymbol{h}$ is the magnetization exchange field of the ferromagnet and $\boldsymbol{\sigma}$ is the Pauli vector. 
This is used as an input in the Usadel equation for the matrix $\hat{h}$, presented in Eqn.~(\ref{Eqn:h}).

\subsection{Spin accumulation without equilibrium contribution}\label{App:Noneq_fieldeff}
We show the non-equilibrium spin accumulation in Fig.~\ref{fig:NonEqSpinAcc}, which governs the voltage-dependence of the total spin accumulation shown in the main body of the manuscript. 
\begin{figure*}
\includegraphics[width=\textwidth, angle=0,clip]{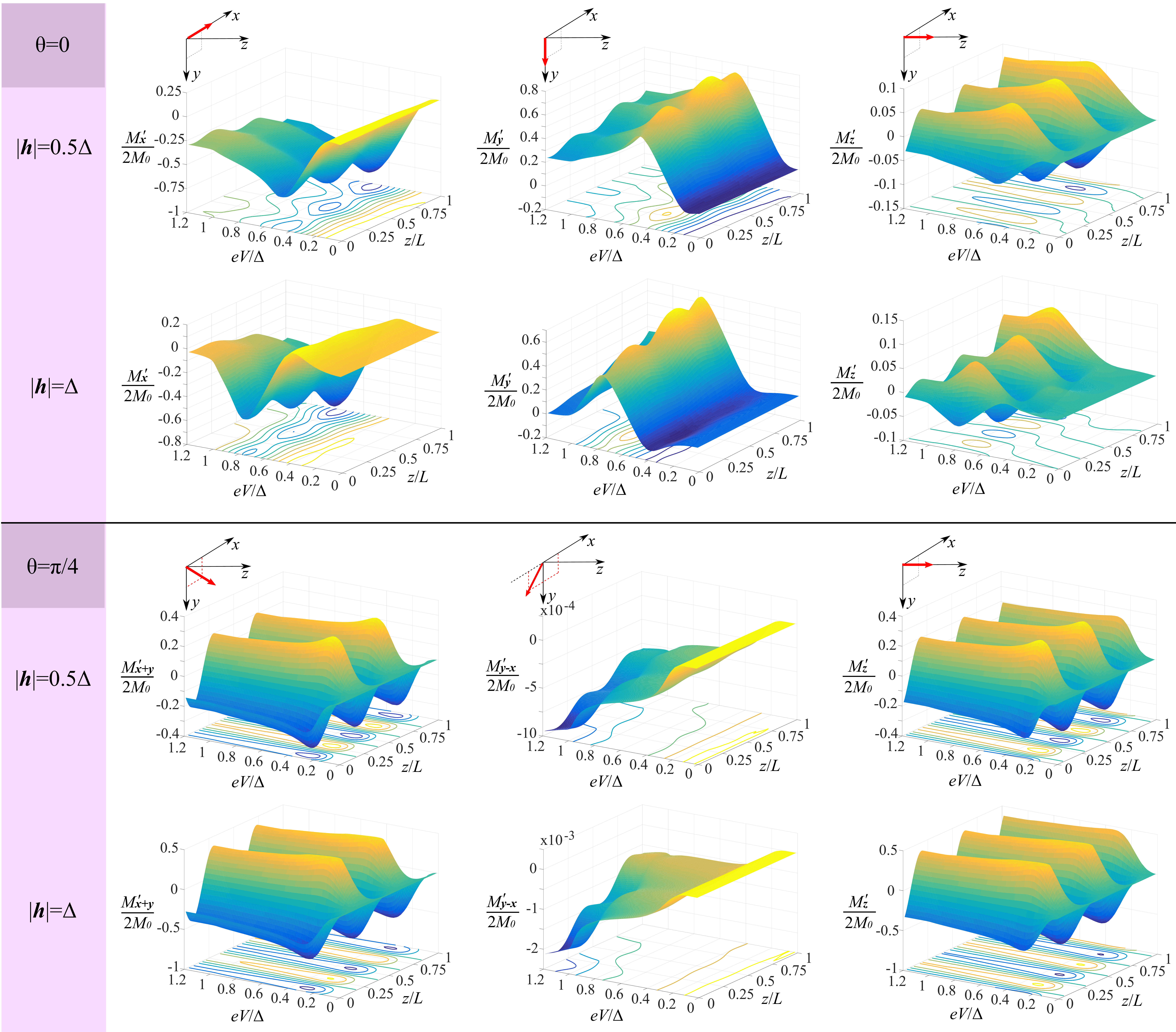}
\caption{(Color online) Spin accumulation $M'_\tau$, without contribution from the equilibrium component of the Green's function, along unit vectors $\tau$ for a nanowire with SOC and Zeeman splitting as a function of the applied voltage bias. All system parameters are as in Fig.~\ref{fig:SpinAcc}.}
\label{fig:NonEqSpinAcc}
\end{figure*}

\subsection{Weak proximity effect solution}\label{App:WPE}
The solution to the weak proximity equations (\ref{Eqn:WPE}) for $\theta=0$ give
\begin{eqnarray}\label{Eqn:0}
f_s&=& C_1\exp\left(-z\sqrt{-2\left(\sqrt{-2iD_F^2\alpha^4-|\boldsymbol{h}|^2}+D_F\alpha^2(i-1)+i\epsilon\right)/D_F}\right)\nonumber\\
&&+C_2\exp\left(z\sqrt{-2\left(\sqrt{-2iD_F^2\alpha^4-|\boldsymbol{h}|^2}+D_F\alpha^2(i-1)+i\epsilon\right)/D_F}\right)\nonumber\\
&&+C_3\exp\left(-z\sqrt{-2\left(\sqrt{-2iD_F^2\alpha^4-|\boldsymbol{h}|^2}-D_F\alpha^2(i-1)-i\epsilon\right)/D_F}\right)\nonumber\\
&&+C_4\exp\left(z\sqrt{-2\left(\sqrt{-2iD_F^2\alpha^4-|\boldsymbol{h}|^2}-D_F\alpha^2(i-1)-i\epsilon\right)/D_F}\right)\nonumber\\
&=& C_1\exp(-zk_1)+C_2\exp(zk_1)+C_3\exp(-zk_2)+C_4\exp(zk_2),\nonumber\\
&=& f_{s:k1}+f_{s:k2},\nonumber\\
d_\parallel &=& \left(D_F\alpha^2(i+1)-i\sqrt{-2iD_F^2\alpha^4-|\boldsymbol{h}|^2}\right)f_{s:k1}/|\boldsymbol{h}| +  \left(D_F\alpha^2(i+1)+i\sqrt{-2iD_F^2\alpha^4-|\boldsymbol{h}|^2}\right)f_{s:k2}/|\boldsymbol{h}|,\nonumber\\
d_\perp &=& C_5\exp\left(z\sqrt{(4D_F \alpha^2(i+1)-2i\epsilon)/D_F}\right)+C_6\exp\left(-z\sqrt{(4D_F \alpha^2(i+1)-2i\epsilon)/D_F}\right).
\end{eqnarray}
Here the notation $f_{s:k1}$ denotes the terms in the solution to the singlet component that have wavevector $k_1$ (\ie the terms with constant prefactors $C_1$ and $C_2$), while $f_{s:k2}$ denotes those with wavevector $k_2$ (\ie with constants $C_3$ and $C_4$). The constant prefactors $C_1-C_6$ can be determined by supplementing the solution with the associated Kupriyanov-Lukichev boundary conditions.

The solution to the weak proximity equations (\ref{Eqn:WPE}) for $\theta=\pi/4$ are rather more complicated:
\begin{eqnarray}\label{Eqn:pi4}
f_s&=& \frac{2D_F\alpha^2}{hq_1}\left(-E_1q_2\exp(-q_3 z)/2 - E_2q_2\exp(q_3 z)/2\right.\nonumber\\
&& \left. -E_3q_4\exp(-q_5 z)/2 - E_4q_4\exp(q_5 z)/2\right.\nonumber\\
&& \left. +iq_6(E_5\exp(-q_7 z) + E_6\exp(q_7 z)\right),\nonumber\\
&=& f_{s:q3}+f_{s:q5}+f_{s:q7},\nonumber\\
d_\parallel &=& \frac{hq_1}{D_F\alpha^2}\left(f_{s:q3}/q_4 - q_{s:q5}/q_2+f_{s:q7}/2iq_6\right),\nonumber\\
d_\perp &=& \frac{h}{2D_F\alpha^2}\left(f_{s:q3}q_9/q_2 + f_{s:q5}q_8/q_4 + f_{s:q7}q_{10}/q_6\right).\nonumber\\
\end{eqnarray}
Here $f_{s:q3}$ contains the terms with wavevector $q_3$ (with constants $E_1$ and $E_2$), $f_{s:q5}$ contains terms with $q_5$ (\ie $E_3$ and $E_4$), and $f_{s:q7}$ contains terms with $q_7$ (\ie $E_5$ and $E_6$). These constants can again be determined by supplementing the solution with the associated Kupriyanov-Lukichev boundary conditions. The functions $q_{1,2,4,6,8,9,10}$ are shorthand for non-exponential functions of the parameters $\alpha, D_F$ and $h$, while functions $q_{3,5,7}$ which appear in the exponentials are also functions of $\epsilon$. As these contain the relevant information about the wavevector, we provide these explicitly:
\begin{eqnarray}
q_3&=&-\left((\sqrt{3}i-1)(8D_F^2\alpha^4+3|\boldsymbol{h}|^2)-2Y^{1/3}(4D_F\alpha^2-3i\epsilon)+Y^{2/3}(\sqrt{3}i+1)\right)^{1/2}/\sqrt{3D_FY^{1/3}},\nonumber\\
q_5&=&-\left((\sqrt{3}i+1)(8D_F^2\alpha^4+3|\boldsymbol{h}|^2)-2Y^{1/3}(4D_F\alpha^2-3i\epsilon)+Y^{2/3}(\sqrt{3}i-1)\right)^{1/2}/\sqrt{3D_FY^{1/3}},\nonumber\\
q_7&=&\sqrt{2}\left(8D_F^2\alpha^4+3|\boldsymbol{h}|^2-Y^{1/3}(4D_F\alpha^2-3i\epsilon)+Y^{2/3}\right)^{1/2}/\sqrt{3D_FY^{1/3}},\nonumber\\
\end{eqnarray}
where $Y=-8D_F^3\alpha^6+9|\boldsymbol{h}|^2D_F\alpha^2+3\sqrt{768D_F^2\alpha^{12}-96D_F^4\alpha^8|\boldsymbol{h}|^2+33D_F\alpha^4|\boldsymbol{h}|^4+3|\boldsymbol{h}|^6}$.

\subsection{$d$-vector and density of states}\label{App:dvec}
Fig.~\ref{fig:dvec} presents the real components of the $d$-vector for field strength $|\boldsymbol{h}|=0.5\Delta$ and rotations $\theta=0$ and $\pi/4$, calculated from the numerical solution to the full Usadel equations.
\begin{figure*}
\includegraphics[width=0.98\textwidth, angle=0,clip]{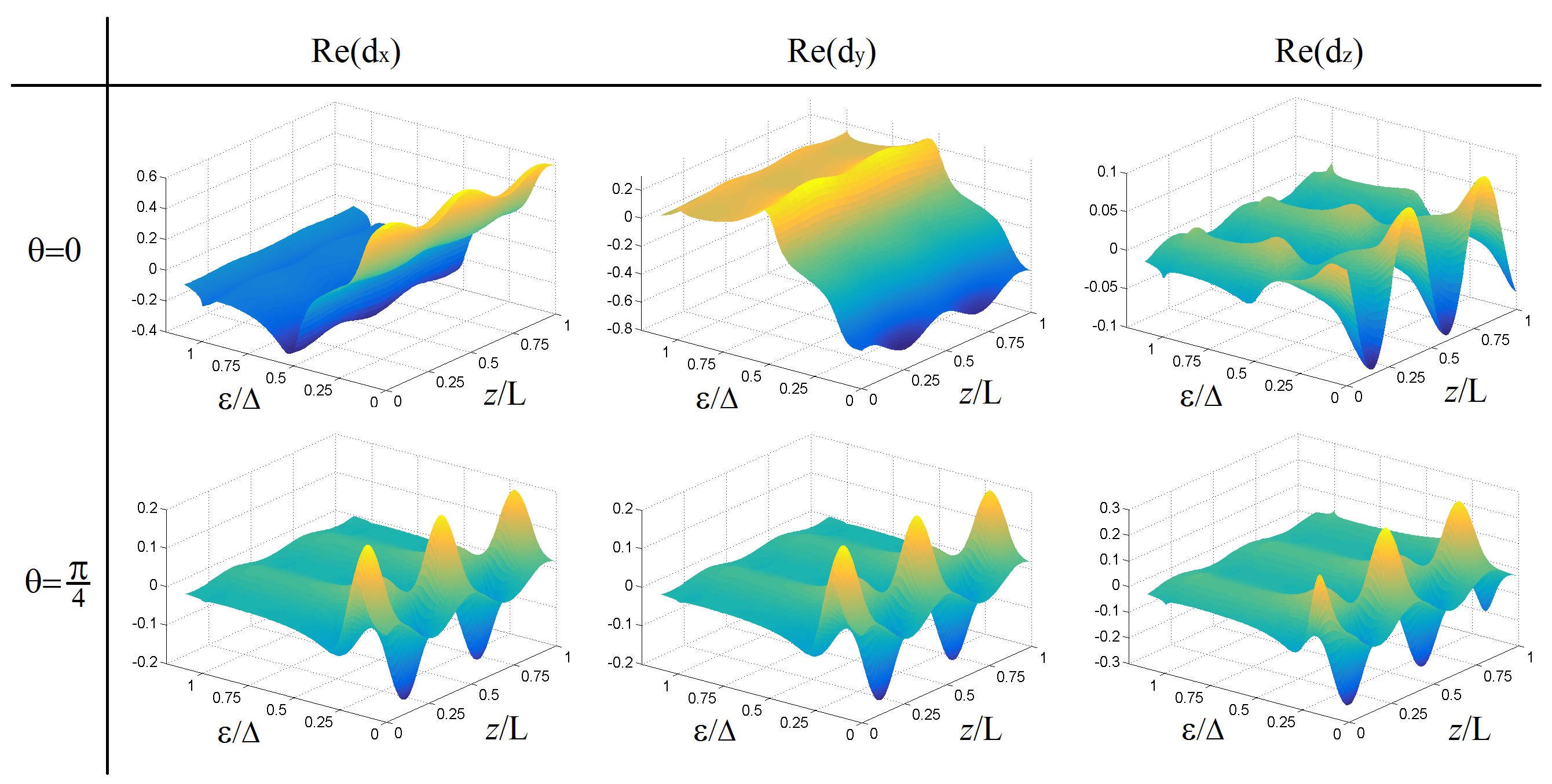}
\caption{(Color online) Components of the $d$-vector at zero bias, for exchange field orientations $\theta=0$ and $\pi/4$. The field strength is $|\boldsymbol{h}|=0.5\Delta$, and similarly all other parameters are as given in Fig.~\ref{fig:SpinAcc}.}
\label{fig:dvec}
\end{figure*}

Fig.~\ref{fig:ldos} shows the local density of states $D(\epsilon)$ in the middle of the sample for field strengths $|\boldsymbol{h}|=0.5\Delta$ and $|\boldsymbol{h}|=\Delta$, and rotations $\theta=0$ and $\theta=\pi/4$.
\begin{figure*}
\includegraphics[width=0.5\textwidth, angle=0,clip]{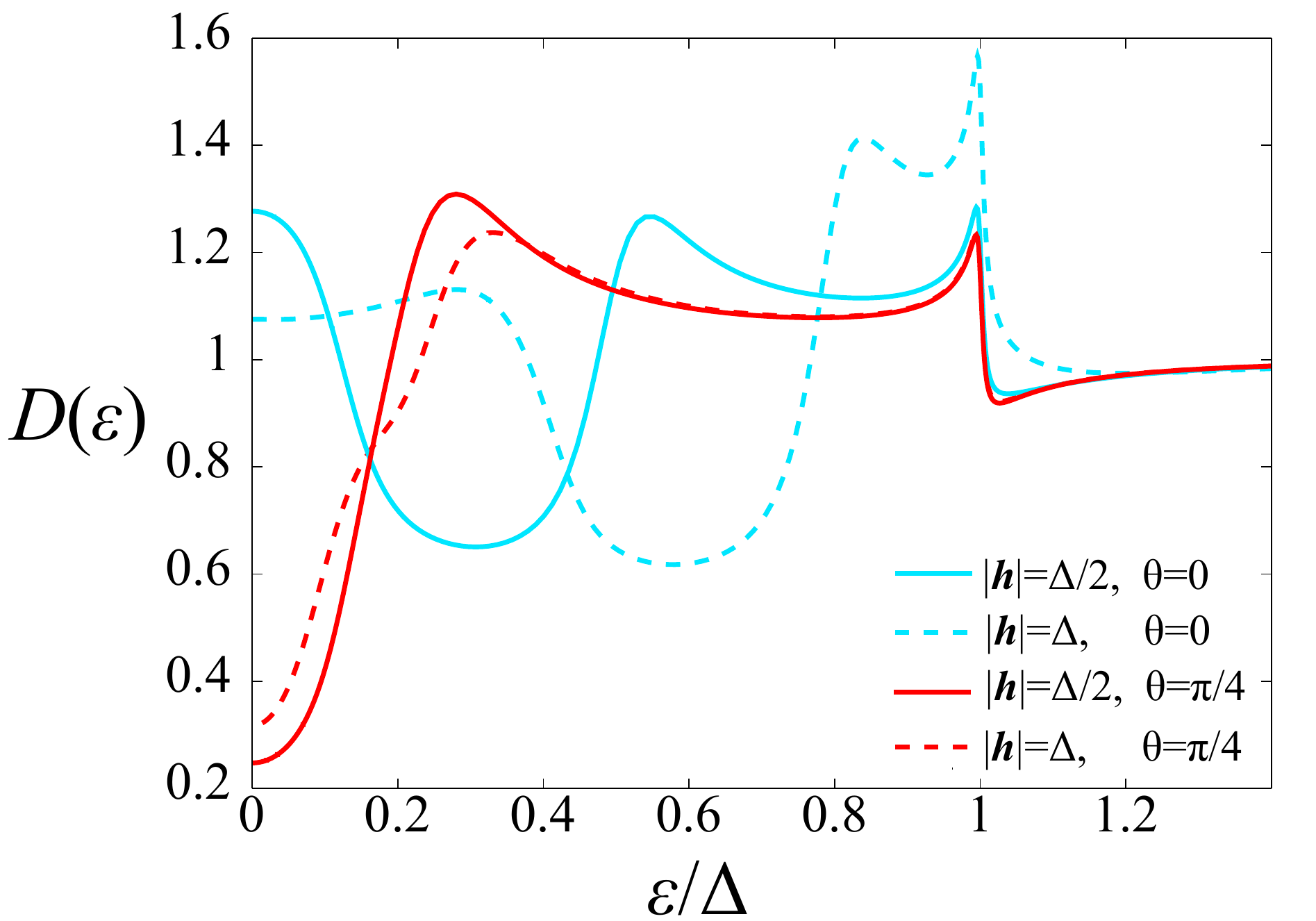}
\caption{(Color online) Local density of states $D(\epsilon)$ in the middle of the sample for field strengths $|\boldsymbol{h}|=0.5\Delta$ and $|\boldsymbol{h}|=\Delta$, and rotations $\theta=0$ and $\theta=\pi/4$.}
\label{fig:ldos}
\end{figure*}

\subsection{Exchange field along wire $\boldsymbol{h}=0.5\Delta\hat{\boldsymbol{z}}$}\label{App:hz}
In Fig.~\ref{fig:hz} we plot the total spin accumulation $M_\tau$ and isolated non-equilibrium component $M'_\tau$ along unit vector $\tau$, for an exchange field aligned along the nanowire, $\boldsymbol{h}=0.5\Delta\hat{\boldsymbol{z}}$. There is near equivalence in the $\hat{\boldsymbol{x}}$ and $\hat{\boldsymbol{y}}$ directions of the spin accumulation, and once again we see a similar influence of the non-equilibrium portion which suppresses the equilibrium contribution and introduces offset oscillations along the wire in the intermediate bias regime. In this case, increasing the field strength increases the magnitude of the spin accumulation, as was the case for exchange field perpendicular to the SO gauge field, and there is no bias-shift in peak magnetization (not shown).
\begin{figure*}
\includegraphics[width=0.98\textwidth, angle=0,clip]{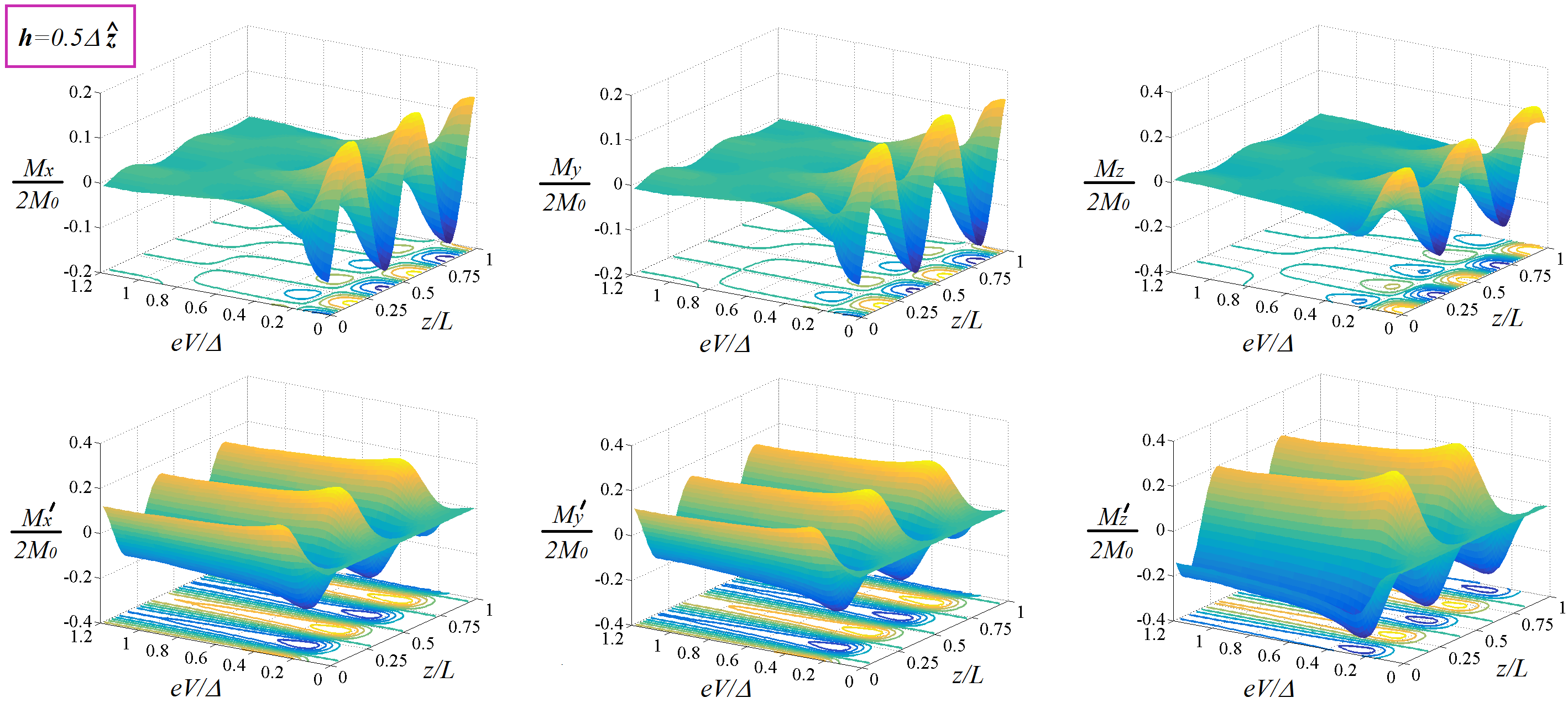}
\caption{(Color online) Total spin accumulation $M_\tau$ and isolated non-equilibrium component $M'_\tau$ along unit vector $\tau$, for exchange field aligned along the nanowire, $\boldsymbol{h}=0.5\Delta\hat{\boldsymbol{z}}$, \ie parallel with the SO gauge field.}
\label{fig:hz}
\end{figure*}

\end{widetext}

\end{document}